\documentclass{elsart1p}
\usepackage{amsmath,amssymb}
\usepackage{epsfig,graphicx}

\usepackage{amssymb}

\usepackage{textcomp}
\usepackage[latin1]{inputenc}
\usepackage{pstricks,pst-node,pst-text,pst-3d}
\usepackage{amsmath}

\newcommand{\IG}[2]{\includegraphics*[#1]{#2}}

\def\FIG #1 #2 [#3] #4\par{%
 \begin{figure}[!h] \begin{center}%
   \includegraphics*[#3]{#2}%
    \\ \parbox{10cm}{%
    \caption{\label{#1}#4}}%
 \end{center}\end{figure}%
}

\begin{document}

\begin{frontmatter}

\title{FSI and rare B decays: $B \to \pi\pi,\rho\rho$}

\author{M.I.Vysotsky}

\address{ITEP, Moscow}

\begin{abstract}
The final state interactions (FSI) model, in which soft
rescattering of low mass intermediate states dominate, is
suggested. It explains why the strong interaction phases are large
in the $B_d\to\pi\pi$ channel and are considerably smaller in the
$B_d\to\rho\rho$ one.

\end{abstract}
\end{frontmatter}

\section{Introduction}

Understanding of final state interactions (FSI) in B decays is
needed in order: 1. to predict/explain the ratios of branching
ratios, $B\to \pi\pi, B\to \rho\rho$ is a very spectacular
example; 2. to study strong interactions; 3. to understand DCPV:
$C \sim \sin\alpha \sin\delta$, so: to understand values of $C$,
$B \to K \pi$ is a very spectacular example.

$C$-averaged branching ratios of $B\to\pi\pi$ and $B\to\rho\rho$
decays are presented in the following Table \cite{1}:

\bigskip

\begin{center}
\begin{tabular}{|c|c|c|c|}
\hline Mode & ${\rm Br}(10^{-6})$ &  Mode & ${\rm Br}(10^{-6})$ \\
\hline $B_d \to\pi^+ \pi^-$ & $5.2 \pm 0.2$ & $B_d \to
\rho^+\rho^-$ & $24 \pm 3$ \\ $B_d \to\pi^0 \pi^0$ & $1.5 \pm
0.2$ & $B_d \to \rho^0\rho^0$ & $0.74 \pm 0.29$ \\ $B_u \to\pi^+
\pi^0$ & $5.6 \pm 0.4$ & $B_u \to \rho^+\rho^0$ & $18.2 \pm 3.0$
\\ \hline
\end{tabular}

\end{center}

\bigskip

where we observe the absence of color suppression (naive factor
$1/3^2/2=1/18$ in decay probability) of $\pi^0\pi^0$ mode.

Charmless strangeless $B$-decays are described by the sum of tree
(T) and penguin (P) Feynman diagrams:

\bigskip

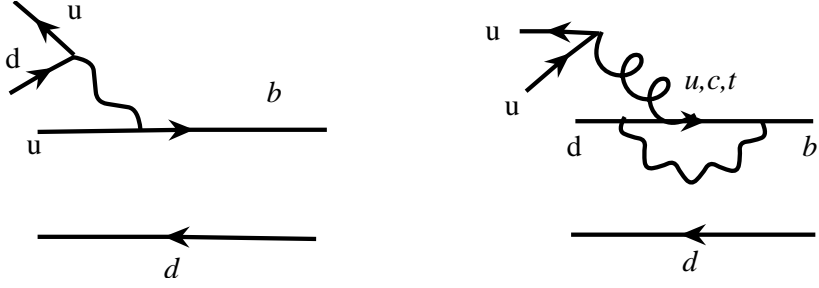
\IG{width=0.8\textwidth}{fig13.epsi}

\bigskip

We work in the so-called ``t-convention'' for penguin amplitudes,
when  $(V_{ub} V_{ud}^* + V_{cb} V_{cd}^* + V_{tb}
V_{td}^*)f(m_c/M_W) =0$ is subtracted from decay amplitudes. In
this convention CKM phases difference of $T$ and $P$ amplitudes is
$\alpha \approx 90^o$ that is why they do not interfere in
C-averaged decay probabilities.

\section{Analysis of experimental data}

Using isotopic invariance of strong interactions $B\to\pi\pi$
decay amplitudes may be presented in the following form:

$$
M_{\bar B_d \to \pi^+\pi^-} = e^{-i\gamma}\frac{1}{2\sqrt 3}A_2
e^{i\delta_2^\pi} +  e^{-i\gamma}\frac{1}{\sqrt 6} A_0
e^{i\delta_0^\pi} + \left|\frac{V_{td}^* V_{tb}}{V_{ub}
V_{ud}^*}\right| e^{i\beta} P e^{i(\delta_P^\pi +
\tilde\delta_0^\pi)} \;\; ,
$$
$$
M_{\bar B_d \to \pi^0\pi^0} = e^{-i\gamma}\frac{1}{\sqrt 3}A_2
e^{i\delta_2^\pi} -  e^{-i\gamma} \frac{1}{\sqrt 6} A_0
e^{i\delta_0^\pi} - \left|\frac{V_{td}^* V_{tb}}{V_{ub}
V_{ud}^*}\right| e^{i\beta} P e^{i(\delta_P^\pi  +
\tilde\delta_0^\pi)} \;\;  ,
$$
$$
M_{\bar B_u \to \pi^-\pi^0} = \frac{\sqrt 3}{2\sqrt 2}
e^{-i\gamma} A_2 e^{i\delta_2^\pi} \;\; .
$$

Neglecting P from 3 branching ratios 3 parameters $A_0, A_2$ and
$|\delta_0 - \delta_2|$ can be extracted, and for phase difference
of the amplitudes with $I = 0$ and 2 we get:
\begin{equation}
\cos(\delta_0^\pi -\delta_2^\pi) = \frac{\sqrt 3}{4} \frac{{\rm
B}_{+-} - 2 B_{00} + \frac{2}{3} \frac{\tau_0}{\tau_+}
B_{+0}}{\sqrt{\frac{\tau_0}{\tau_+} B_{+0}}\sqrt{B_{+-} + B_{00}
-\frac{2}{3} \frac{\tau_0}{\tau_+} B_{+0}}} \nonumber\;\; .
\end{equation}
Using experimentally measured branching ratios from the Table we
obtain: $|\delta_0^\pi - \delta_2^\pi| = 55^o$.

$P^2$ term  we can extract from Br$(K^0\pi^+)$:
Br$(B_d\to\pi^+\pi^-)_P \approx 0.59\cdot10^{-6}$; subtracting it
from experimental data we see that penguin amplitude diminishes a
bit phase difference: $|\delta_0^\pi - \delta_2^\pi| =
47^o\pm10^o$. Let us remind that in the case of $D \to \pi\pi$
decays phase difference of isotopic amplitudes is two times larger
\cite{2}: $|\delta_0^D - \delta_2^D| = 86^o \pm 4^o$, which
suggest approximate $1/M$ scaling of FSI phases, where $M$ is the
mass of decaying meson.

$\rho$-mesons produced in $B$-decays are almost completely
longitudinally polarized, so the analysis goes just like for
$\pi$-mesons: $|\delta_0^{\rho} - \delta_2^{\rho}| =
15^{o}+5^o-10^o$, small (unlike pion case).

This difference of  FSI phases  is responsible for different patterns of
$B\to\pi\pi$ and $B\to\rho\rho$ decay probabilities.

We want to understand why FSI phases are large  in  $B\to\pi\pi$
amplitudes but small in $B\to\rho\rho$ amplitudes.

pQCD: PHASES  ARRIVE FROM LOOPS, SO THEY ARE SMALL, which is
correct for $B\to\rho\rho$-decays but it does not work for
$B\to\pi\pi$.

SO: DYNAMICS at LARGE DISTANCES MATTER.

\section{Model of FSI}

Which intermediate states are important (here we follow papers
\cite{3}, \cite{4}, see also \cite{5})?

$b \to u \bar{u} d$ decay produce mostly 3 isotropically oriented
jets of light mesons, each having about 1.5 GeV energy. In $e^+
e^-$ annihilation at 3 GeV c.m. energy average charged particles
(pions) multiplicity is about 4 - so, taking $\pi^0$'s into
account in B-mesons decays to light quarks  in average 9 ``pions''
are produced, flying in 3 widely separated directions (or almost
isotropically, taken transverse momentum into account). Branching
ratio of such decays is large, about $10^{-2}$. However such
states {\it{NEVER}} rescatter into two pions or two $\rho$-
mesons.

Which intermediate states will transform into two mesons final
state we easily understand  studying  inverse process of two light
mesons scattering at 5 GeV c.m. energy. In this process two jets
of particles moving in the directions of initial particles are
formed. Energy of each jet is $M_B/2$, while its invariant mass
squared is not more than $M_B \Lambda_{QCD}$.

Following these arguments in the calculation of the imaginary
parts of decay amplitudes we will take two particle intermediate
states into account, to which branching ratios of $B$-mesons are
maximal:

\bigskip

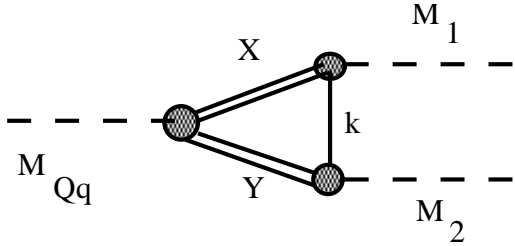
\IG{width=0.5\textwidth}{fig1_1.epsi}

\bigskip

It is convenient to transform integral over longitudinal
(relatively to outgoing meson momentum) and time-like components
of $k$ in the following way:
$$
\int dk_0dk_z=1/(2 \cdot M_B^2)\int ds_X ds_Y \;\; .
$$

Integrals over $s$ rapidly decrease when $s$ increase since only
low mass clusters contribute to amplitude of 2 meson production.
In this way we get:
$$
M_{\pi\pi}^I = M_{XY}^{(0)I} (\delta_{\pi X}\delta_{\pi Y} + i
T_{XY \to\pi\pi}^{J=0}) \;\;.
$$

Since Br$B\to\rho\rho$ is large it contributes a lot to FSI phase
of $B\to \pi\pi$ decay; NOT VICE VERSA! $B\to\rho\rho\to\pi\pi$
chain can be calculated with the help of unitarity relation; for
small $t$ we can trust elementary $\pi$-meson exchange in $t$-
channel:
$$
{\rm Im} M(B\to\pi\pi) = \int\frac{d\cos\theta}{32\pi}
M(\rho\rho\to\pi\pi) M^* (B\to\rho\rho) \;\; .
$$

Introducing formfactor $exp(t/\mu^2)$ for $\mu^2=2m_\rho^2$ we
obtain: $\delta_0^\pi(\rho\rho) = 15^o \; , \delta_2^\pi(\rho\rho)
= -5^o \;\; , \delta_0^\pi(\rho\rho) - \delta_2^\pi(\rho\rho) =
20^o$ and half of experimentally observed phase difference is
explained. Let us emphasize that  $\delta_I^\pi(\rho\rho) \sim
1/M_B \rightarrow 0$.

It is remarkable that FSI phases generated by
$B\to\pi\pi\to\rho\rho$ chain are damped by ${\rm
Br}(B\to\rho^+\rho^-, \rho^+\rho^0)/{\rm Br}(B\to\pi^+\pi,
\pi^+\pi^0)$ ratios and are a few degrees: $\delta_0^\rho(\pi\pi)
- \delta_2^\rho(\pi\pi) \approx 4^o $.

For $\pi\pi$ intermediate state we take Regge model expression for
$T_{\pi\pi\to\pi\pi}$, which takes into account pomeron, $\rho$
and $f$ trajectories exchange. Pomeron exchange produces imaginary
T and does not contribute to phase shifts as far as it is
critical, $\alpha_P(0) = 1$. However, for the amplitude of the
supercritical pomeron exchange we have: $T\sim (s/s_0)^{
\alpha_P(t)}(1+exp(-i\pi\alpha_P(t)))/(-\sin(\pi\alpha_P(t)))=(s/s_0)^
{(1+\Delta)}(i+\Delta\pi/2)$, where in the last expression $t=0$
was substituted and the value of intercept
$\alpha_P(0)=1+\Delta,\Delta\approx0.1$ was used. So,
$\delta_0^\pi(\pi\pi) = 5.0^o \; , \;\delta_2^\pi(\pi\pi) = 0^o$.

$\pi a_1$ intermediate state should also be taken into account.
Large branching ratio of $B_d\to\pi^{\pm} a_1^{\mp}$-decay (${\rm
Br}(B_d\to\pi^{\pm} a_1^{\mp})=(32 \pm 4)*10^{-6}$) is partially
compensated by small $\rho\pi a_1$ coupling constant (it is $1/3$
of $\rho\pi \pi$ one): $\delta_0^\pi(\pi a_1) = 4^o \; ,
\;\delta_2^\pi(\pi a_1) = -2^o$, where we assume that the sign of
$\pi a_1 $ contribution to phases difference is the same as that
of the elastic channel.

Finally: $\delta_0^\pi = 23^o \; , \;\;\delta_2^\pi =
-7^o\;\;,\;\; \delta_0^\pi -\delta_2^\pi = 30^o $, and the
accuracy of this number is not high.

In conclusion the model of FSI in $B\to M_1 M_2$ decays is
suggested; it explains the absence of color suppression of $B \to
\pi^0\pi^0$ decay. Relatively small $B \to \pi^+\pi^-$ branching
ratio is the reason why $B \to \rho^0\rho^0$ mode remains small.

$B \to\pi\pi$: we cannot reproduce $C_{+-}$ value measured by
Belle (-0.55(9)) while BABAR result (-0.25(8)) is much more
acceptable and we predict almost maximal DCPV in $B(\bar B) \to
\pi^0\pi^0$ decays: $C_{00} \approx -0.60$ .

I am very grateful to the organizers of PANIC 08 conference for
the great time in Israel and to RFBR for the travel grant RFBR
08-02-08858.

\end{document}